\documentclass[12pt,letterpaper]{article}
\usepackage[T1]{fontenc}
\usepackage{tgheros}

\usepackage{inconsolata}
\usepackage{microtype}
\usepackage{amsmath}
\usepackage[helvet]{sfmath}
\usepackage[super,sort,comma]{natbib}
\usepackage[margin=1in,headheight=15pt,headsep=15pt]{geometry}
\usepackage{setspace}
\usepackage{ragged2e}
\RaggedRight
\usepackage{fancyhdr}
\usepackage[section]{placeins}
\usepackage{graphicx}
\usepackage{float}
\usepackage[font={small,sf},labelfont=bf,justification=raggedright,singlelinecheck=false]{caption}
\newcommand{\captionv}[3]{\begin{center}\parbox{#1cm}{\caption[#2]{#3}}\end{center}}
\usepackage{etoolbox}
\AtBeginEnvironment{table}{\singlespacing}
\makeatletter
\renewcommand\@biblabel[1]{#1.}
\makeatother
\usepackage{xcolor}
\usepackage{xurl}
\usepackage[hidelinks]{hyperref}
\hypersetup{pdfauthor={Xianghuan He, Changyu Sun, Andy N. Wiles, Yao Hao, Gregory B. Biedermann, Brian W. Pogue, Rongxiao Zhang},pdfsubject={},pdfkeywords={},pdftitle={From Radiotherapy Dataset to Cherenkov Dose Map: A Digital Linac Monte Carlo Simulation Pipeline}}
\usepackage[all]{hypcap}
\renewenvironment{abstract}{\section*{Abstract}\noindent}{\par}
\begin{document}
\color{black}
\pagenumbering{arabic}
\begin{center}
{\Large\bfseries From Radiotherapy Dataset to Cherenkov Dose Map: A Digital Linac Monte Carlo Simulation Pipeline\par}

\medskip
{\normalsize
Xianghuan He\textsuperscript{1}, Changyu Sun\textsuperscript{1,2},
Andy N. Wiles\textsuperscript{3}, Yao Hao\textsuperscript{4},\\[3pt]
Gregory B. Biedermann\textsuperscript{3}, Brian W. Pogue\textsuperscript{5},
Rongxiao Zhang\textsuperscript{3,*}\par}
\medskip
{\small
\textsuperscript{1}Department of Chemical and Biomedical Engineering, College of Engineering,
University of Missouri, Columbia, Missouri, United States.\par\smallskip
\textsuperscript{2}Department of Radiology, School of Medicine,
University of Missouri, Columbia, Missouri, United States.\par\smallskip
\textsuperscript{3}Department of Radiation Oncology, School of Medicine,
University of Missouri, Columbia, Missouri, United States.\par\smallskip
\textsuperscript{4}Division of Medical Physics, Department of Radiation Oncology,
Washington University School of Medicine, St. Louis, Missouri, United States.\par\smallskip
\textsuperscript{5}Thayer School of Engineering, Dartmouth College,
Hanover, New Hampshire, United States.\par\medskip
\textsuperscript{*}Corresponding author: Rongxiao Zhang,
\href{mailto:rongxiao.zhang@health.missouri.edu}{rongxiao.zhang@health.missouri.edu}\par}
\end{center}

\begin{abstract}
\noindent {\bf Background:} Cherenkov-based radiotherapy (RT) monitoring relies on a quantitative link between Cherenkov emission and dose.
That link spans local photon creation along charged-particle tracks, optical transport in tissue, escape at the surface, and camera detection.
Its most upstream stage, the local correspondence between energy deposition and Cherenkov creation at the same position, cannot be measured directly and  therefore requires paired dose and Cherenkov creation records for characterization.\\
{\bf Purpose:} We hypothesized that a Geant4-based digital linear accelerator (linac) framework could generate, from DICOM RT input, Monte Carlo (MC) dose records paired with local Cherenkov creation records, allowing the local dose--Cherenkov relationship to be quantified under matched delivery and geometry conditions.
Specifically, we assessed whether the framework could translate DICOM RT datasets into paired  dose and Cherenkov records, whether the  dose component reproduced TPS-computed RTDOSE (RD) spatial patterns, and whether the paired Cherenkov records retained dose-related spatial information.\\
{\bf Methods:} A Geant4-based digital linac framework was developed for  MC computation of dose and Cherenkov in a water-equivalent phantom.
It converts a DICOM RT dataset consisting of CT, structure set, plan, and dose objects into executable jaw, MLC, gantry, and collimator states, then scores energy deposition and Cherenkov photon creation positions on  a common voxel grid with 2.5~mm isotropic spacing within the same transport history.
Static jaw-defined fields, beam-quality cases, and a two-arc volumetric modulated arc therapy (VMAT) plan were evaluated.
The  dose distribution was compared with TPS-computed RD using Acuros XB (AXB) and a secondary Anisotropic Analytical Algorithm (AAA) reference.
Local  dose--Cherenkov association and normalized spatial-shape agreement were characterized at voxel, plane, and profile levels.\\
{\bf Results:}  RD--Dose full-volume 3D gamma passing rates (3\%/3 mm, 10\% threshold) were 97.05\% to 98.86\% for the AXB-referenced  field-size and beam-quality cases (97.47\% with the AAA reference) and 97.84\% for the VMAT composite.
Local Cherenkov creation remained strongly associated with the paired  dose: 3D voxel \(r^2\) was at least 0.9890, and projection and  profile \(r^2\) at least 0.9942, in every case.
Local shape agreement was case dependent:  Dose--Cherenkov 3D gamma was 97.97\% to 99.16\% for the \(10\times10\) 6X, 10X, and 15X fields and 96.59\% for the VMAT composite, but 71.22\% for \(5\times5\) and 32.82\% for 6FFF.
Profile and penumbra analyses localized the \(5\times5\) difference along depth and near the field edges; the two-edge penumbra fraction reached 16.5\% versus 4.3\% for \(20\times20\), while the 6FFF difference was predominantly depth oriented.\\
{\bf Conclusions:} In a homogeneous water phantom, local Cherenkov creation follows local dose closely within the field interior, while systematic differences appear near beam edges and under FFF beam quality, before any optical transport is introduced.
The framework provides paired records for characterizing this local relationship and lays the foundation to investigate Cherenkov transport with tissue optics, prediction of surface-emitted signals, and correlation with superficial dose measurements for patient-specific RT.\\

\end{abstract}

\newpage

\section{Introduction}

\noindent
Cherenkov emission is produced during megavoltage external-beam radiotherapy (RT) when charged particles set in motion by the beam exceed the phase velocity of light in the local medium.
The emission originates in the same charged-particle transport that deposits dose, and it has been applied to treatment verification, delivery monitoring, and superficial dose estimation.\cite{Helo2014CerenkovQA,Alexander2021ColorCherenkov,Hachadorian2020BreastCherenkov,Decker2024CIEDCherenkov,Okamoto2026LinacCherenkovQA,Chen2025CherenkovPositioning,Geiersbach2026CherenkovSurfaceMapping}
The link between emission and dose spans four stages: local photon creation along charged-particle tracks, optical transport in tissue, escape at the surface, and camera detection.\cite{Ciarrocchi2017CerenkovPhysics}
The most upstream stage is the local correspondence between energy deposition and Cherenkov creation at the same position.
It is not directly measurable, and characterizing it therefore requires a paired data generation and comparison  workflow driven by patient-specific RT plan input.
\par

This correspondence has been examined previously under defined conditions.
Glaser et al.\cite{Glaser2014CherenkovDose} compared Cherenkov emission with absorbed dose for photon, electron, and proton beams using theory and Monte Carlo (MC) simulation, and reported a linear relationship above threshold energies together with residual depth-dependent differences that decreased as field size increased.
Zlateva et al.\cite{Zlateva2019Formalism,Zlateva2019ElectronBeam} established a formalism for in-water Cherenkov emission-based dosimetry and computed MC Cherenkov-to-dose conversion factors for photon and electron beams.
Zhang et al.\cite{Zhang2013SuperficialCherenkov} related Cherenkov emission to superficial dose for megavoltage photon beams, with the largest differences near field edges.
Zhu et al.\cite{Zhu2026CherenkovTSETOptical} reported that emission per unit dose in total-skin electron therapy is a function of tissue optical properties.
Parks et al.\cite{Parks2025HybridCherenkov} coupled a treatment planning system (TPS) dose calculation with an optical MC stage in a hybrid model evaluated in simple phantom geometry.
These studies show that the correspondence is quantifiable and that it varies with beam energy, field size, depth, and medium.
They evaluate it for predefined beam configurations, or from a TPS-computed dose distribution, rather than systematically generating dose and Cherenkov together from the RT plan input.
\par

In parallel, the components needed to generate dose and Cherenkov records from RT plans have been developed separately.
Geant4\cite{Agostinelli2003Geant4} and TOPAS\cite{Faddegon2020TOPAS} provide particle transport with optical physics, and the fidelity of simulated Cherenkov quantities has been assessed directly.
Trigila et al.\cite{Trigila2022Geant4Cherenkov} showed that Cherenkov photon accuracy in a Geant4-based GATE model depends on the parameterization of primary-electron propagation, Khodaei et al.\cite{Khodaei2024TOPASOptical} evaluated TOPAS for Cherenkov generation, optical-photon transport, and radioluminescence-based dosimetry, and Zlateva et al.\cite{Zlateva2020StepSize} reported the step-size dependence of calculated Cherenkov-to-dose conversion factors.
Plan-driven MC execution has been demonstrated in parallel.
Ortiz et al.\cite{Ortiz2024TPS2TOPAS} parameterized TrueBeam treatment plans from DICOM RT input into TOPAS control files, and Lee et al.\cite{Lee2019GATEVMAT} converted volumetric modulated arc therapy (VMAT) plan parameters for TrueBeam with an HD120 multileaf collimator (MLC) into GATE dose calculations for pretreatment quality assurance.
These studies score dose rather than examine its link to Cherenkov emission: Cherenkov creation is not retained from the same transport history, and the delivery provenance that associates each record with  the delivery state under which it was generated is not preserved.
Bringing the two capabilities into a single transport configuration is desirable, for the local  dose--Cherenkov comparison as well as optical transport and emission processes in patient-specific geometries.

\begin{figure}[!htbp]
\centering
\includegraphics[width=0.90\textwidth]{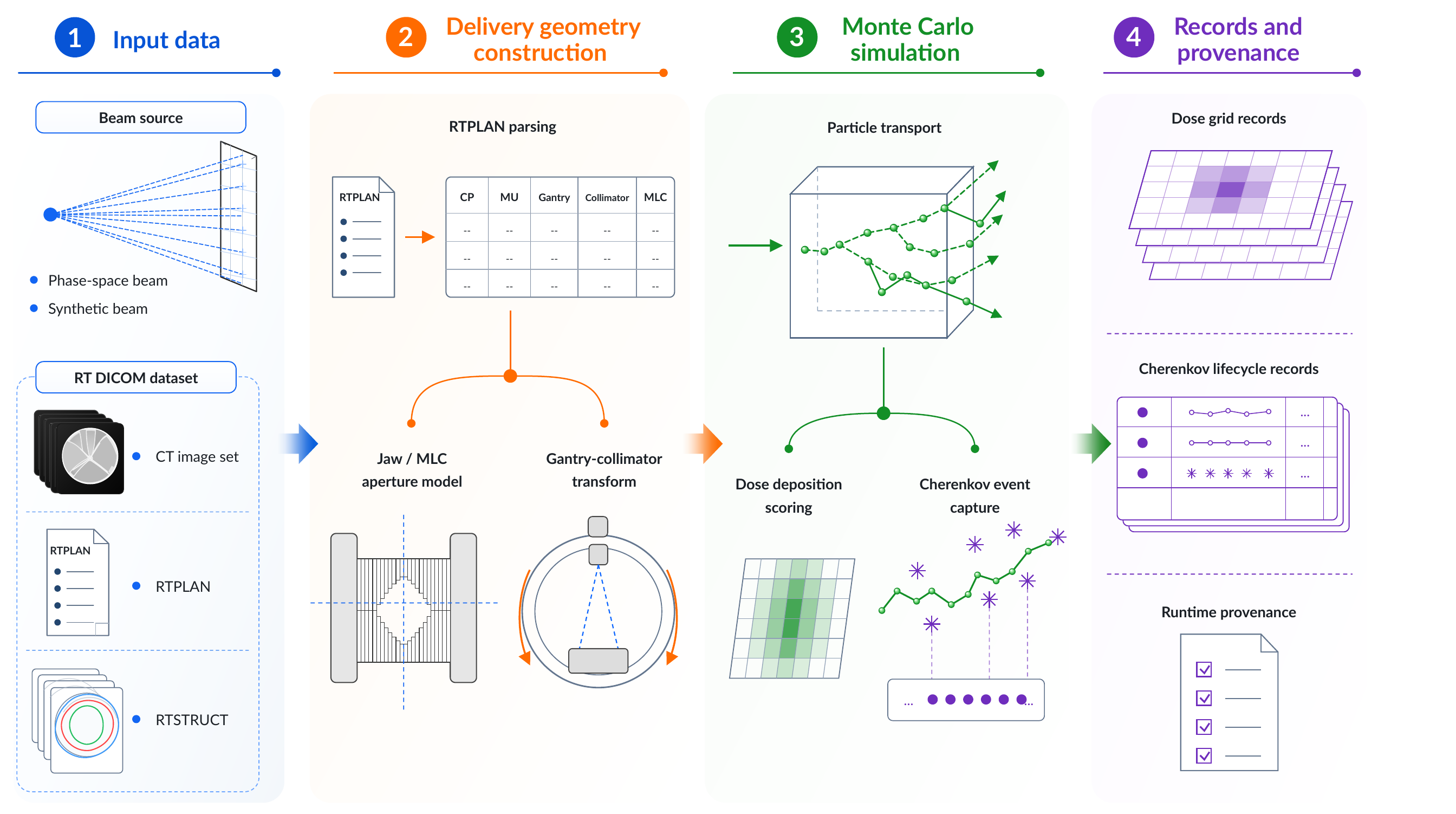}
\captionv{15}{DICOM RT-driven paired-record generation}{ Overview of the DICOM RT-driven framework for generating paired  dose and local Cherenkov-creation records.
The four stages are input data, delivery geometry construction from RTPLAN control point (CP) information, Monte Carlo simulation, and records and provenance.
Beam-source data, the computed tomography (CT)-derived phantom, and RTPLAN define each simulated case.

Energy deposition and Cherenkov photon creation positions are scored on a common voxel grid within the same transport history.
TPS-computed RTDOSE remains external to particle transport and is used only as the reference for  RD--Dose comparison.
\label{fig:overview}}
\end{figure}

In this work, we use a Geant4-based digital linac framework to generate paired records of  dose and local Cherenkov creation from DICOM RT datasets using MC simulations.
The framework converts the DICOM RT Plan (RTPLAN) within each dataset into executable jaw, MLC, gantry, and collimator states, transports particles through a water-equivalent phantom, and scores energy deposition and Cherenkov photon creation on a common voxel grid within the same transport history.
In the scope of this manuscript, a Cherenkov record denotes simulation-derived local creation information rather than a measured signal; photon-level attributes are retained, and the voxel-level creation counts used here are a reduction of them.
Figure~\ref{fig:overview} summarizes the workflow in four stages, from input data through delivery-geometry construction and MC transport to paired outputs, with TPS-exported DICOM RT Dose (RTDOSE; denoted RD in the comparisons below) placed outside particle transport and entering only as an external reference.
Evaluation then proceeds in two steps.
The  dose component is first compared with RTDOSE for the same plan, so that a subsequent  Dose--Cherenkov difference can be interpreted in the context of the local  dose--Cherenkov relationship rather than attributed to delivery reconstruction.
Local dose--Cherenkov agreement is then quantified across field size, photon beam quality, and modulated delivery.
\par

\section{Materials and Methods}

\subsection{\texorpdfstring{DICOM RT-driven generation of paired  dose and local Cherenkov records}{DICOM RT-driven generation of paired dose and local Cherenkov records}}

The simulation begins with precomputed TrueBeam phase-space particle records sampled at the source boundary.
The electron gun and other accelerator components upstream of this boundary are not explicitly re-simulated.
Case-specific delivery is controlled by the RTPLAN-defined jaws, HD120 multileaf collimator (MLC), and gantry and collimator orientations before the particles are transported into the CT-derived phantom.

As shown in Figure~\ref{fig:overview}, the first stage assembles the input data.
Each DICOM RT dataset comprised CT, RT Structure Set (RTSTRUCT), RT Plan (RTPLAN), and RT Dose (RD) objects for the same planned delivery.
CT and RTSTRUCT defined the voxelized phantom and scoring space.
The phase-space beam source supplies the incident particle records, and RTPLAN provides the planned delivery description.

The second stage converts each RTPLAN description into executable delivery geometry using its jaw openings, MLC leaf positions, gantry and collimator poses, control-point states, and delivery weights, following the DICOM RT object model and International Electrotechnical Commission (IEC) coordinate convention.\cite{DICOM2026cPS33,IEC61217_2011}
The IEC-based delivery geometry was transformed into a simulation world coordinate system aligned with the DICOM patient coordinates. For the fixed-field cases, the beam propagated along the positive Y direction.
To examine the local dose--Cherenkov relationship under simple and modulated delivery conditions, we prepared static fixed-field plans and a two-arc VMAT plan. Specifically, these fixed-field plans were used to compare different field sizes at a fixed beam quality and different beam qualities at a fixed field size.
The fixed-field plans used static jaw-defined square apertures, whereas  the VMAT plan was divided into delivery segments between adjacent RTPLAN control points. Each segment was simulated using the fixed geometry of its starting control point. The resulting dose and Cherenkov creation-count distributions were accumulated on a common phantom grid within each arc and then summed across the two arcs.\cite{Otto2008VMAT,Kim2014VMATControlPoints}

\noindent

The third stage executes Monte Carlo transport using the constructed delivery geometry, beam source, phantom, material assignment, and physics settings.
For each constructed delivery state, incident particle records from the selected TrueBeam phase-space file were used as primary inputs and transported through the RTPLAN-derived delivery geometry and CT-derived phantom using Geant4.

Simulations used Geant4 version 11.3.2 with the \texttt{FTFP\_BERT} modular physics list, with its electromagnetic (EM) component replaced by \texttt{G4EmStandardPhysics\_option3}.
A secondary-production range cut of 0.01~mm was used, matching the 0.01~mm range cut documented in the headers of the TrueBeam phase-space files used in this study.
The default ionisation \texttt{StepFunction} for electrons and positrons in this physics constructor was retained, with \texttt{dRoverRange} = 0.2 and \texttt{finalRange} = 0.1~mm.
\texttt{G4OpticalPhysics} was registered to generate Cherenkov photons within the same transport histories.
The default \texttt{G4Cerenkov} step-limiting parameters were retained: a maximum average photon number per step of 100 and a maximum relative change in $\beta=v/c$ per step of 10\%.\cite{Geant4ApplicationDevelopersOptical,Geant4Source1132}

The dose scoring grid matches the spacing, dimensions, and position of the corresponding RD grid. Energy deposition is accumulated voxelwise across transport histories, using 2.5~mm isotropic voxels.

The fourth stage stores the generated data and their associated provenance.
The accumulated energy-deposition values are stored voxelwise as the  dose record.
For each Cherenkov photon, the stored attributes can include its creation position and direction, final position and direction, final energy, Geant4 event identifier, and track identifier.
For the present analyses, the photon creation positions are binned onto the  dose voxel grid to form the paired local Cherenkov creation-count distribution.
Runtime configurations, delivery manifests, output metadata, and analysis provenance are stored alongside these physical records.

\subsection{\texorpdfstring{Simulation and Analysis}{Simulation and Analysis}}

To characterize the local dose--Cherenkov relationship, we evaluated the conditions summarized in Table~\ref{tab:simulation_conditions}.

The VMAT plan comprised one clockwise and one counterclockwise arc, each containing 180 RTPLAN control points that defined 179 delivery segments.
The number of phase-space particle records simulated for each delivery segment was determined by the planned delivery weights, without monitor-chamber feedback.
The two arcs were assigned 52.35\% and 47.65\% of the total records, respectively, and their resulting distributions were summed directly.
The field-size series replayed the same 6X phase-space source, whereas the beam-quality series used the corresponding 6X, 6 MV flattening-filter-free (FFF, labeled 6FFF), 10X, and 15X phase-space sources; each replay used more than \(2.1\times10^{9}\) phase-space particle records.

The external dose references were generated with Acuros XB (AXB) and, for the secondary comparison, the Anisotropic Analytical Algorithm (AAA).

\begin{table}[H]
\centering
\scriptsize
\captionv{15}{Evaluated simulation conditions and analysis series}{Evaluated simulation conditions and analysis series.
In the Modality column, field-size entries denote  fixed jaw-defined square fields, whereas ``Two-arc VMAT'' denotes the TPS-defined MLC-modulated VMAT plan.
\label{tab:simulation_conditions}}
\resizebox{\textwidth}{!}{
\begin{tabular}{llll}
\hline
Energy & Modality & RD reference & Analysis series \\
\hline
6X & \(5\times5\)~cm\(^2\) & AXB & Field size \\
6X & \(10\times10\)~cm\(^2\) & AXB; AAA & Field-size / beam-quality baseline \\
6X & \(20\times20\)~cm\(^2\) & AXB & Field size \\
10X & \(10\times10\)~cm\(^2\) & AXB & Beam quality \\
15X & \(10\times10\)~cm\(^2\) & AXB & Beam quality \\
6FFF & \(10\times10\)~cm\(^2\) & AXB & Beam quality \\
6X & Two-arc VMAT & AXB & Modulated delivery \\
\hline
\end{tabular}
}
\end{table}

The evaluated cases were constructed from TPS-exported DICOM RT datasets (Varian ARIA and Eclipse 16.1.4).
Two Eclipse photon dose-calculation algorithms were used to generate the external RD references.
Of the two Eclipse algorithms, AAA is a convolution/superposition algorithm, whereas AXB deterministically solves the linear Boltzmann transport equation for coupled photon and electron transport.\cite{Bragg2006AAA,Hoffmann2012AcurosXB}
Because MC simulation describes particle transport through a stochastic solution of the same governing transport equation, the transport formulation of AXB is methodologically closer to MC than the analytical formulation of AAA.
AXB RD was therefore used as the primary external reference for all evaluated cases, while an additional AAA RD was retained only for the 6X \(10\times10\)~cm\(^2\) case as a reference-algorithm secondary check.
The CT-derived phantom spanned \(51.0\times51.0\times51.75\)~cm\(^3\).
CT and RTSTRUCT defined a water-equivalent Body region with air outside the Body contour on the native \(512\times512\times207\) CT grid, which had approximately \(0.996\times0.996\times2.5\)~mm voxel spacing.
MC energy deposition and Cherenkov creation positions were scored on the corresponding RD-matched grids with 2.5~mm spacing: \(203\times203\times201\) voxels for most AXB  field-size and beam-quality cases, \(202\times202\times201\) voxels for the AAA check, and \(202\times201\times201\) voxels for the VMAT arc grids.

The analysis  involved two comparisons.
First, the  dose component was compared with the corresponding external RD  to validate the framework's reliability in reproducing the TPS-computed spatial dose pattern under the same planned delivery.
Second, the paired  dose and local Cherenkov creation-count distributions were analyzed to quantify their voxelwise association and normalized spatial-shape agreement. Spatial agreement was assessed over the full volume and further examined in representative planes and profiles.

Analyses were performed on normalized 3D distributions.
For the  field-size and beam-quality cases, RD defined the central-axis depth of maximum dose, \(d_{\mathrm{max}}\), and the RD,  dose, and Cherenkov volumes were each divided by their own mean value within the same  spherical region of interest (ROI) with a radius of 0.5~cm centered at that position. The choice of \(d_{\mathrm{max}}\) as the normalization depth follows the relative dose--Cherenkov comparison of Glaser et al.\cite{Glaser2014CherenkovDose}
For the final VMAT composite, the RD global 3D maximum defined a common  spherical ROI with a radius of 0.5~cm, and the RD,  dose, and Cherenkov volumes were each divided by their own mean over the same ROI voxels.\par
 Full-volume 3D gamma analysis used a global 3\%/3~mm criterion and a 10\% threshold  relative to the maximum of the normalized reference volume, with points below this threshold excluded from the passing-rate calculation.\cite{Low1998Gamma,Miften2018TG218}\par

For  RD--Dose gamma, RD was the reference and  dose was the evaluation distribution; for  Dose--Cherenkov gamma,  dose was the reference and the Cherenkov creation-count distribution was the evaluation distribution.
The  RD--Dose calculation was used to check the  dose-producing chain, whereas  Dose--Cherenkov gamma was interpreted as a relative spatial-shape comparison rather than clinical dose validation.\cite{Ma2020TG157,Geurts2022MPPG5b}
We use \(P_{\gamma}^{(3)}\) for the independently calculated full-volume 3D passing rate and \(P_{\gamma}^{(2)}\) for an independently calculated representative-plane 2D passing rate; the latter is not a slice of the 3D gamma result.

 To complement the spatial agreement analysis, voxelwise association between paired  dose, \(D_{\mathrm{MC}}\), and local Cherenkov creation intensity, \(C_{\mathrm{Ch}}\), was quantified using the squared Pearson correlation coefficient

\begin{equation}
r^2=\left[\mathrm{corr}\left(D_{\mathrm{MC}},C_{\mathrm{Ch}}\right)\right]^2.
\label{eq:squared_pearson}
\end{equation}

 For the field-size series, beam-quality series, and VMAT composite, correlation was evaluated over paired positions for which the corresponding  dose representation was at least 10\% of its maximum, and was reported for full 3D voxels, the X--Z projection, and  one-dimensional profiles. These were central-window-averaged depth profiles for the jaw-defined cases and Y-direction profiles for the VMAT composite.

 Representative-plane maps were extracted as single selected layers of the three-dimensional distributions. Their 2D gamma maps, together with one-dimensional profiles, were used to examine local spatial agreement and identify regions of discrepancy.
For the jaw-defined cases, the transverse X--Z plane was extracted at the RD-defined \(d_{\mathrm{max}}\), and the central X--Y plane was evaluated at \(Z=0\).\par For representative-plane 2D gamma analysis, the same 3\%/3~mm criterion was applied, with thresholds defined relative to the maximum of each reference plane. A 10\% threshold was used for the field-size and beam-quality series. In contrast, a 40\% threshold was used for the representative plane of the VMAT composite. This higher threshold excluded the spatially dispersed low-value regions, allowing the planar gamma analysis to focus on spatial agreement within the compact central high-value region.\par For the jaw-defined cases, central-axis percentage depth dose (PDD) profiles for RD and  dose, the corresponding central-axis Cherenkov profile, and lateral X profiles at \(d_{\mathrm{max}}\) were compared.
Difference profiles were calculated as RD minus  dose and Cherenkov creation counts minus  dose.
To quantify the relative lateral-edge contribution, the  dose lateral profile was used to calculate the 50\% field width \(W_{50}\), the mean 80\%--20\% penumbra width \(\bar p_{80-20}\), and the two-edge penumbra fraction

\begin{subequations}\label{eq:edge_metrics}
\begin{align}
W_{50} &= x_{50,R}-x_{50,L},
\label{eq:w50}\\
\bar p_{80-20} &= \frac{p_L+p_R}{2},
\label{eq:mean_penumbra}\\
f_{\mathrm{edge}} &= \frac{p_L+p_R}{W_{50}}\times100\%.
\label{eq:edge_fraction}
\end{align}
\end{subequations}

where \(x_{50,L}\) and \(x_{50,R}\) are the left and right 50\% crossing positions, and \(p_L\) and \(p_R\) are the corresponding 80\%--20\% edge widths obtained by linear interpolation between adjacent grid samples.
For the control-point diagnostic, one representative control-point-defined delivery state was selected from each arc of the VMAT RTPLAN.
Each selected state was executed as an independent Geant4 simulation using its RTPLAN-defined MLC leaf positions, jaw openings, and collimator geometry, the complete merged 6X phase-space source, and a common gantry-zero diagnostic orientation.
The two standalone simulations were analyzed separately at the control-point level, while the weighted outputs from both arcs were merged into the final VMAT composite for the plan-level full-volume gamma analysis.

\section{\texorpdfstring{Results}{Results}}

Across the evaluated delivery conditions, paired  dose and local Cherenkov-creation records retained a strong overall spatial association, whereas their local spatial-shape agreement varied substantially among cases.
The results below first quantify this overall association and then examine the case-dependent local spatial agreement.

\subsection{\texorpdfstring{Local Cherenkov creation remained spatially associated with paired  dose}{Local Cherenkov creation remained spatially associated with paired dose}}\label{sec:results_correlation}

Using the squared Pearson correlation defined in Eq.~(\ref{eq:squared_pearson}), the paired records within the  dose-based in-field mask remained strongly correlated at the 3D voxel, X--Z projection, and  profile levels in every tested case (Table~\ref{tab:cherenkov_full_correlations}).
In the 6X field-size series, the observed 3D voxelwise \(r^2\) values were 0.9915, 0.9933, and 0.9940 for \(5\times5\), \(10\times10\), and \(20\times20\), respectively.
With the aperture fixed at \(10\times10\), the corresponding 6X value was 0.9933, the 10X and 15X values were 0.9960 and 0.9970, and 6FFF produced the lowest observed voxelwise value of 0.9890.
The two-arc VMAT composite remained within the same high-association range, with a 3D voxelwise \(r^2\) of 0.9898 despite the modulated delivery.
After dimensional reduction, X--Z projection \(r^2\) values ranged from  0.9983 to  0.9997 and  profile values ranged from 0.9942 to 0.9988, showing that the broad dose-related spatial trend was retained even where the voxelwise distributions exhibited greater dispersion.

\begin{table}[!htbp]
\centering
\scriptsize
\captionv{15}{ Dose--Cherenkov correlations across evaluated cases}{ Dose--Cherenkov correlations across evaluated cases.
Squared Pearson coefficients compare  dose with local Cherenkov creation intensity at paired positions on the common grid within the  dose-based in-field mask.
Columns report full 3D voxels, an X--Z projection, and  one-dimensional profiles.
The  field-size and beam-quality rows each represent an independent paired simulation record; the VMAT row reports the VMAT composite.
\label{tab:cherenkov_full_correlations}}
\resizebox{\textwidth}{!}{
\begin{tabular}{llccc}
\hline
Energy & Modality & 3D voxel \(r^2\) & X--Z projection \(r^2\) &
 Profile \(r^2\)\\
\hline
6X & \(5\times5\)~cm\(^2\) & 0.9915 & 0.9994 &  0.9953\\
6X & \(10\times10\)~cm\(^2\) & 0.9933 & 0.9992 & 0.9972\\
6X & \(20\times20\)~cm\(^2\) & 0.9940 &  0.9993 &  0.9977\\
10X & \(10\times10\)~cm\(^2\) & 0.9960 &  0.9997 &  0.9985\\
15X & \(10\times10\)~cm\(^2\) & 0.9970 & 0.9996 & 0.9988\\
6FFF & \(10\times10\)~cm\(^2\) & 0.9890 &  0.9983 & 0.9942\\
6X & VMAT composite & 0.9898 & 0.9987 & 0.9976\\
\hline
\end{tabular}
}
\end{table}

\begin{figure}[!htbp]
\centering
\includegraphics[width=0.90\textwidth]{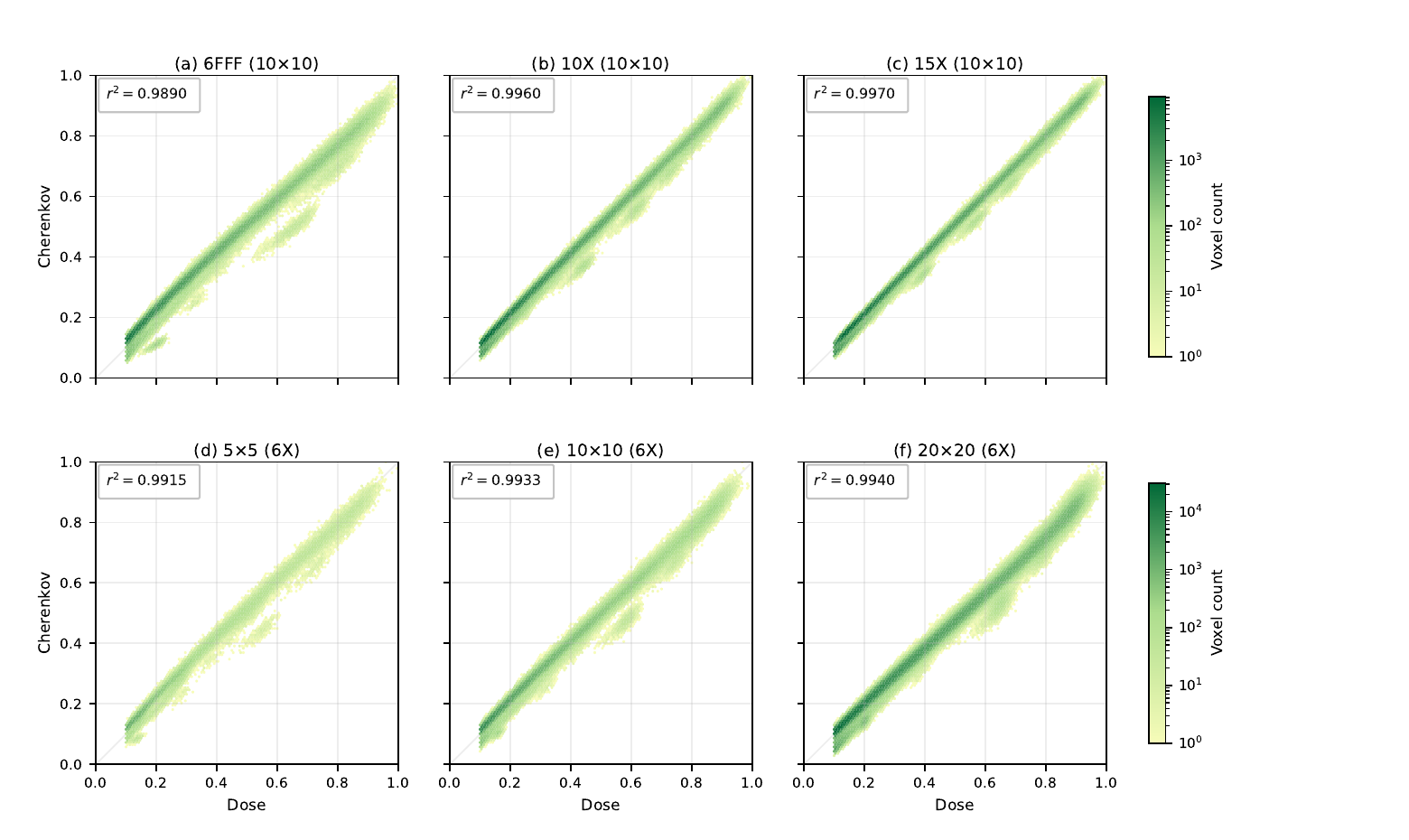}
\captionv{15}{Voxelwise 3D  Dose--Cherenkov correlations}{Voxelwise 3D  Dose--Cherenkov correlations.
Hexbin density plots compare paired  dose and local Cherenkov creation intensity on the common analysis grid, with each axis independently rescaled by the maximum of its corresponding 3D record.
The \(10\times10\) beam-quality comparison spans 6X, 6FFF, 10X, and 15X, with the 6X case shown as the middle panel of the bottom row.
The bottom row fixes the photon beam quality at 6X and compares \(5\times5\), \(10\times10\), and \(20\times20\) fields.
Color indicates voxel count per hexagonal bin on a logarithmic scale, with a separate count scale for each row.
The \(r^2\) values quantify association among the paired voxel values; color intensity and the faint diagonal guide do not define the correlation or agreement with a reference line.
\label{fig:cherenkov_voxel_correlation}}
\end{figure}

The voxelwise distributions in Figure~\ref{fig:cherenkov_voxel_correlation} show the same positive co-variation underlying the high coefficients in Table~\ref{tab:cherenkov_full_correlations}.
Across the 6X field-size series, all three distributions remained concentrated along a common positive trend, consistent with the narrow 0.9915--0.9940 voxelwise \(r^2\) range.
The beam-quality comparison showed greater separation: the 10X and 15X distributions were tightly concentrated, whereas 6FFF exhibited the broadest band and a visible secondary branch, consistent with its lower voxelwise \(r^2\).
Because field size changes the number of in-field voxels and the two rows use separate count scales, differences in color intensity were not interpreted as differences in correlation strength.

\FloatBarrier

Across the evaluated cases, the high \(r^2\) values indicate that greater local energy deposition was generally associated with a greater number of Cherenkov photons created in the same voxel.
This positive association reflects a consistent dose-related trend, but correlation alone does not determine whether the two normalized three-dimensional distributions agree locally in magnitude and position.
We therefore used the full-volume 3D gamma analysis described in Methods to evaluate spatial agreement for both the paired  Dose--Cherenkov records and the  RD and dose distributions.
For  RD--Dose, TPS-computed RTDOSE served as the planning-dose reference for assessing whether the  dose component reproduced the expected spatial dose pattern; for  Dose--Cherenkov, the paired records were compared to determine how closely local Cherenkov creation followed the spatial distribution of  dose.

\subsection{\texorpdfstring{TPS-referenced  RD--Dose consistency and case-dependent  Dose--Cherenkov agreement}{TPS-referenced RD--Dose consistency and case-dependent Dose--Cherenkov agreement}}\label{sec:results_gamma}

The  dose distributions showed high full-volume spatial agreement with the corresponding TPS-exported RD, with  RD--Dose 3D gamma passing rates above 97\% in every tested case (Table~\ref{tab:validation_matrix}).

\begin{table}[!htbp]
\centering
\scriptsize
\captionv{15}{Full-volume 3D gamma summary for  RD--Dose and  Dose--Cherenkov comparisons}{Full-volume 3D gamma summary for  RD--Dose and  Dose--Cherenkov comparisons.
The field-size series tests TPS-defined jaw apertures, the beam-quality series tests photon beam qualities under the same fixed aperture, and the VMAT row reports the VMAT composite.

The AAA row provides an additional comparison for the 6X \(10\times10\) case, with dose and Cherenkov creation counts evaluated on the AAA RD-matched grid.
Both gamma columns report full-volume 3D passing rates, \(P_{\gamma}^{(3)}\), on complete co-registered volumes.
 RD--Dose compares the  dose component with TPS-exported RD after the declared normalization and supports the TPS-referenced spatial check of the  dose component;  Dose--Cherenkov compares the paired Cherenkov creation-count distribution with  dose as a relative spatial-shape analysis.
\label{tab:validation_matrix}}
\resizebox{\textwidth}{!}{
\begin{tabular}{lllrr}
\hline
Energy & Modality & Reference & \shortstack[r]{\\RD--Dose\\3D gamma} & \shortstack[r]{\\Dose--Cherenkov\\3D gamma} \\
\hline
6X & \(5\times5\)~cm\(^2\) & AXB RD & 97.05\% & 71.22\% \\
6X & \(10\times10\)~cm\(^2\) & AXB RD & 98.22\% & 97.97\% \\
6X & \(10\times10\)~cm\(^2\) & AAA RD & 97.47\% & 97.82\% \\
6X & \(20\times20\)~cm\(^2\) & AXB RD & 98.86\% & 90.79\% \\
10X & \(10\times10\)~cm\(^2\) & AXB RD & 97.62\% & 98.53\% \\
15X & \(10\times10\)~cm\(^2\) & AXB RD & 98.47\% & 99.16\% \\
6FFF & \(10\times10\)~cm\(^2\) & AXB RD & 98.56\% & 32.82\% \\
6X & VMAT composite & AXB RD & 97.84\% & 96.59\% \\
\hline
\end{tabular}
}
\end{table}

Among the primary AXB-referenced cases, the observed 6X field-size values were 97.05\%, 98.22\%, and 98.86\% for \(5\times5\), \(10\times10\), and \(20\times20\), respectively.
The \(10\times10\) beam-quality series remained within a similarly narrow range, with passing rates of 97.62\%, 98.47\%, and 98.56\% for 10X, 15X, and 6FFF, respectively.
The two-arc VMAT composite produced a passing rate of 97.84\%, which  remained above 97\%.
 For the 6X \(10\times10\) case,  RD--Dose \(P_{\gamma}^{(3)}\) was 98.22\% with the primary AXB RD reference and 97.47\% with the additional AAA RD reference.

The Dose--Cherenkov passing rates were similar for the AXB and AAA comparisons (97.97\% and 97.82\%, respectively). These comparisons used the respective TPS-exported RD grids specified in Section~II.B.
Across changes in field size and photon beam quality, and for the more complex two-arc VMAT delivery, all  RD--Dose \(P_{\gamma}^{(3)}\) values remained above 97\%, showing that the  dose distributions consistently reproduced the three-dimensional spatial patterns of the corresponding TPS-computed RD.
These results provide evidence that the proposed framework reliably translated the evaluated DICOM RT plans into executable delivery geometries and  dose distributions under the tested homogeneous water-phantom conditions.

With the  dose component showing consistently high TPS-referenced spatial agreement, the  Dose--Cherenkov gamma comparison addressed a different question: how closely the normalized Cherenkov-creation distribution followed the local spatial shape of its paired  dose.
In contrast to the narrow  RD--Dose range,  Dose--Cherenkov \(P_{\gamma}^{(3)}\) varied from 32.82\% to 99.16\% across the independent cases.
The \(10\times10\) 6X, 10X, and 15X fields produced high passing rates of 97.97\%, 98.53\%, and 99.16\%, respectively, while the VMAT composite remained at 96.59\%.
The field-size series was nonmonotonic: \(P_{\gamma}^{(3)}\) increased from 71.22\% for \(5\times5\) to 97.97\% for \(10\times10\), but decreased to 90.79\% for \(20\times20\).
The most pronounced difference occurred for the \(10\times10\) 6FFF field, for which \(P_{\gamma}^{(3)}\) was 32.82\%.
The contrast between metrics was clearest for \(5\times5\) and 6FFF: their voxelwise \(r^2\) values remained high at 0.9915 and 0.9890, whereas their  Dose--Cherenkov gamma passing rates decreased to 71.22\% and 32.82\%.
These results indicate that the local dose--Cherenkov relationship was not characterized by a single level of spatial agreement, but varied with field size and photon beam quality despite consistently strong overall co-variation.
 Although the full-volume metrics revealed differences in agreement across cases, they did not directly indicate where these differences occurred or how they varied spatially. We therefore examined representative transverse and longitudinal planes, together with depth and lateral profiles, to characterize these differences in greater detail.

\subsection{\texorpdfstring{Representative-plane and profile analysis of  Dose--Cherenkov agreement}{Representative-plane and profile analysis of Dose--Cherenkov agreement}}
\FloatBarrier
\subsubsection{\texorpdfstring{ Dose--Cherenkov agreement across field sizes and photon beam qualities}{Dose--Cherenkov agreement across field sizes and photon beam qualities}}\label{sec:results_fieldsize_beamquality_planes}
Within the 6X field-size series, full-volume  Dose--Cherenkov \(P_{\gamma}^{(3)}\) was highest for \(10\times10\), lower for \(20\times20\), and lowest for \(5\times5\).
To further investigate the local dose--Cherenkov relationship across these field sizes, we examined representative transverse and longitudinal planes extracted from the co-registered three-dimensional RD,  dose, and Cherenkov distributions, followed by central profile and lateral penumbra analyses.
The RD and  dose panels provided the plane-level dose comparison, supplemented by the plane-based  RD--Dose gamma maps.
The  Dose--Cherenkov maps were then used to describe how local dose--Cherenkov agreement varied across field size and orientation.
\begin{figure}[H]
\centering
\includegraphics[width=0.99\textwidth]{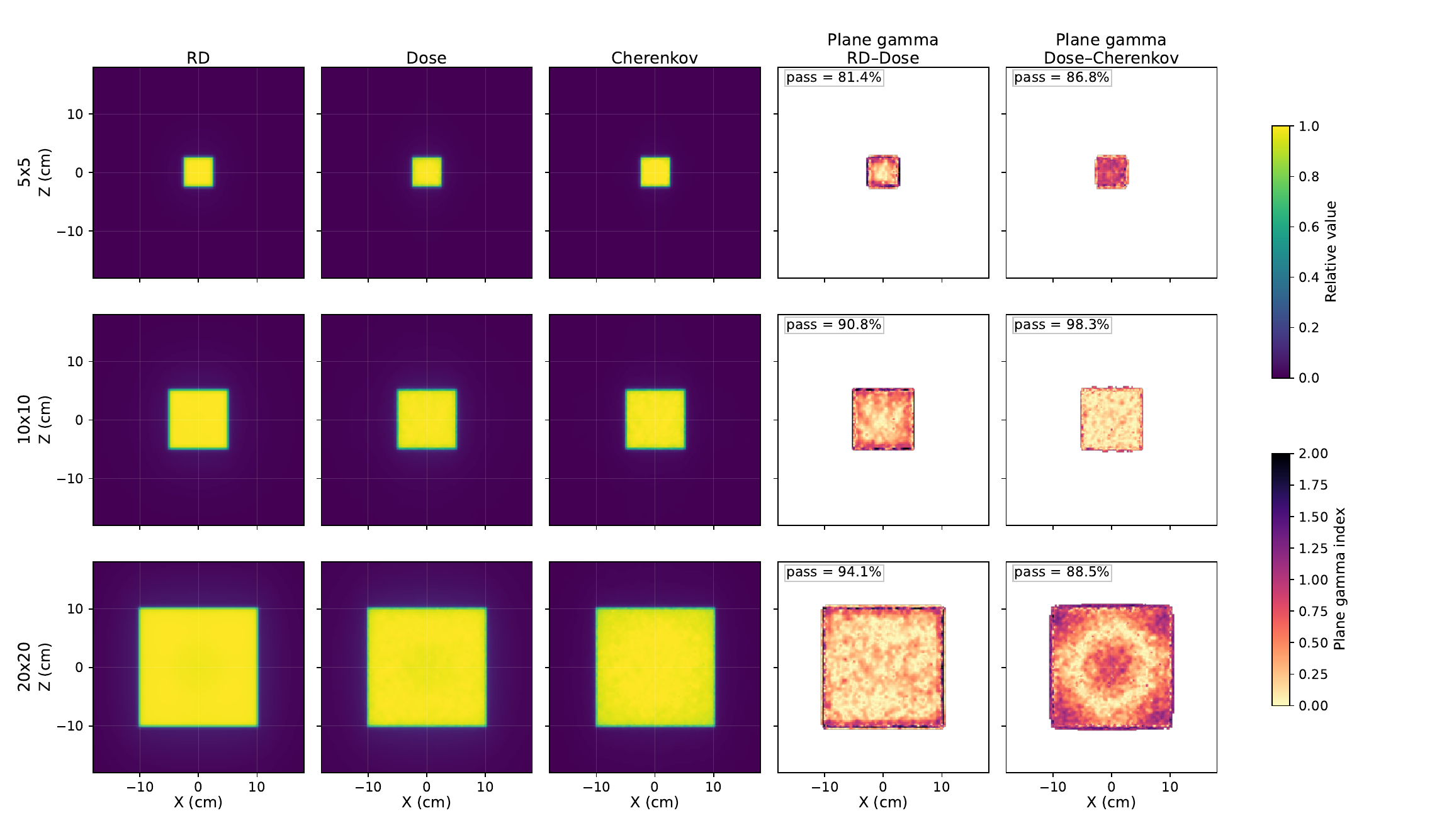}
\captionv{15}{X--Z maps at the RD-defined dmax for three 6X jaw apertures}{X--Z maps at the RD-defined \(d_{\mathrm{max}}\) for three 6X jaw apertures.
Rows show \(5\times5\), \(10\times10\), and \(20\times20\) fields generated from the RTPLAN X and Y jaw settings.
Columns show RD,  dose,  Cherenkov creation counts, the representative-plane 2D  RD--Dose gamma map, and the representative-plane 2D  Dose--Cherenkov gamma map. The relative-value colorbar applies to the first three columns, and the gamma-index colorbar to the last two.
The in-panel pass labels report \(P_{\gamma}^{(2)}\) from these independent 2D calculations; they do not report a slice of the full-volume 3D result.
\label{fig:dose_gamma_fixedfield_xz_cherenkov}}
\end{figure}
Figure~\ref{fig:dose_gamma_fixedfield_xz_cherenkov} shows that the RD,  dose, and Cherenkov maps retained the centered jaw-defined aperture across the three field sizes on the transverse X--Z plane at the RD-defined \(d_{\mathrm{max}}\).
The  RD--Dose gamma maps showed residual structure concentrated near the high-gradient field edges.
The  Dose--Cherenkov results followed a different, nonmonotonic pattern: \(P_{\gamma}^{(2)}\) was 86.8\%, 98.3\%, and 88.5\% for \(5\times5\), \(10\times10\), and \(20\times20\), respectively.
On this transverse plane, the \(10\times10\) field had the closest  Dose--Cherenkov agreement, whereas both the smaller and larger apertures showed more extensive local differences.
\begin{figure}[H]
\centering
\includegraphics[width=0.99\textwidth]{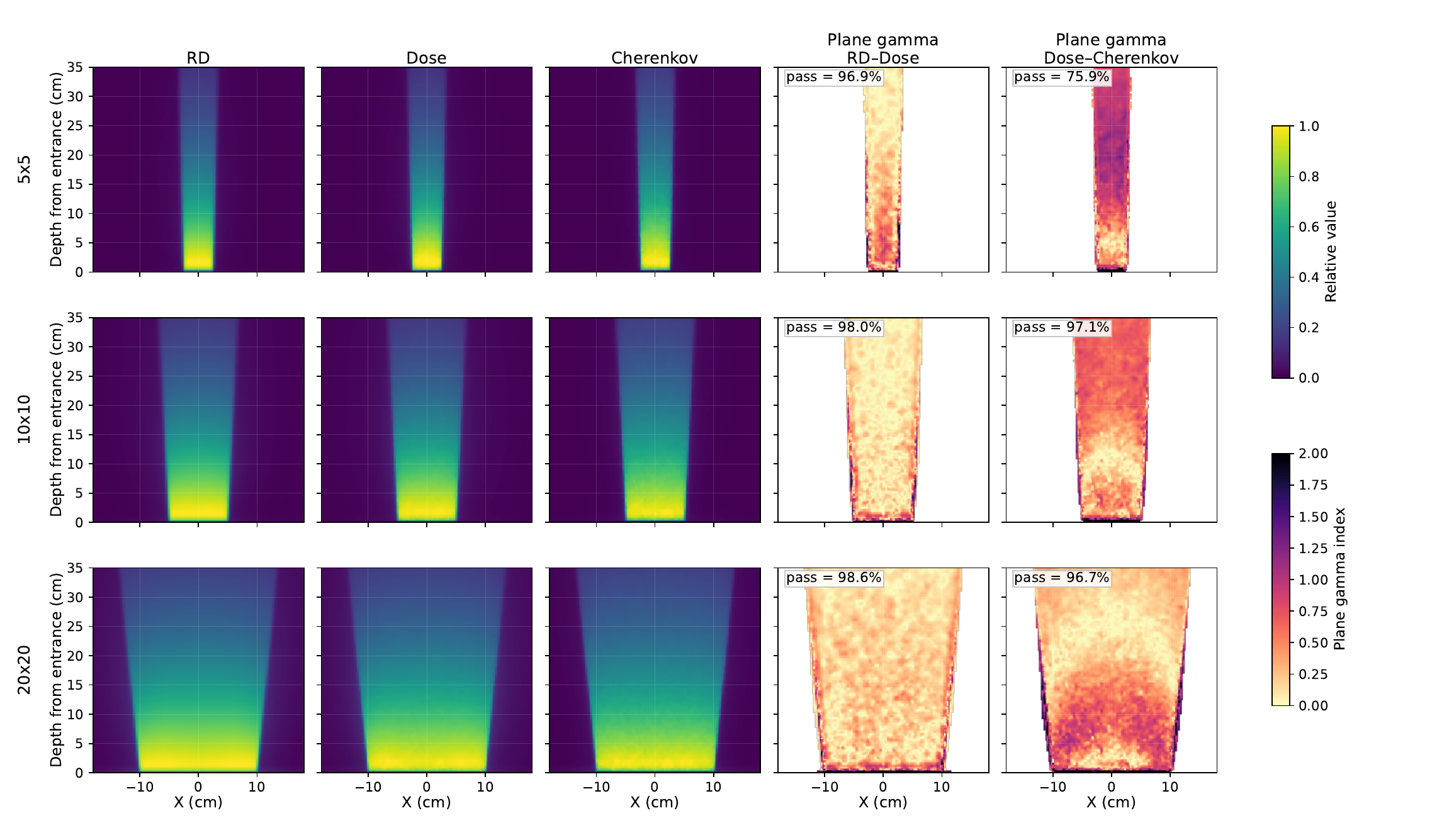}
\captionv{15}{Central X--Y maps for three 6X jaw-defined field sizes}{Central X--Y maps for three 6X jaw-defined field sizes.
Here, Y is the beam-depth direction and is displayed as depth from the phantom entrance.
Rows show \(5\times5\), \(10\times10\), and \(20\times20\) fields on the \(Z=0\) plane.
Columns match Figure~\ref{fig:dose_gamma_fixedfield_xz_cherenkov}: RD,  dose,  Cherenkov creation counts, the representative-plane 2D  RD--Dose gamma map, and the representative-plane 2D  Dose--Cherenkov gamma map.
In-panel pass labels report \(P_{\gamma}^{(2)}\) for this longitudinal plane, where Y is depth.
\label{fig:dose_gamma_fixedfield_xy_cherenkov}}
\end{figure}
Figure~\ref{fig:dose_gamma_fixedfield_xy_cherenkov} shows that the RD and  dose distributions retained similar entrance-to-depth geometry on the longitudinal X--Y plane, with  RD--Dose \(P_{\gamma}^{(2)}\) exceeding 96.9\% for all three field sizes.
The  Dose--Cherenkov values were 75.9\%, 97.1\%, and 96.7\% for \(5\times5\), \(10\times10\), and \(20\times20\), respectively.
For \(5\times5\), the lower full-volume  Dose--Cherenkov agreement was accompanied by reduced agreement on both representative planes.
Within its longitudinal gamma map, gamma-index values generally increased with depth away from the phantom entrance, corresponding to lower local agreement at greater depths.
For \(20\times20\), the lower full-volume agreement was accompanied mainly by the transverse-plane difference, while the longitudinal-plane value remained high at 96.7\%.
\begin{figure}[H]
\centering
\includegraphics[width=0.99\textwidth]{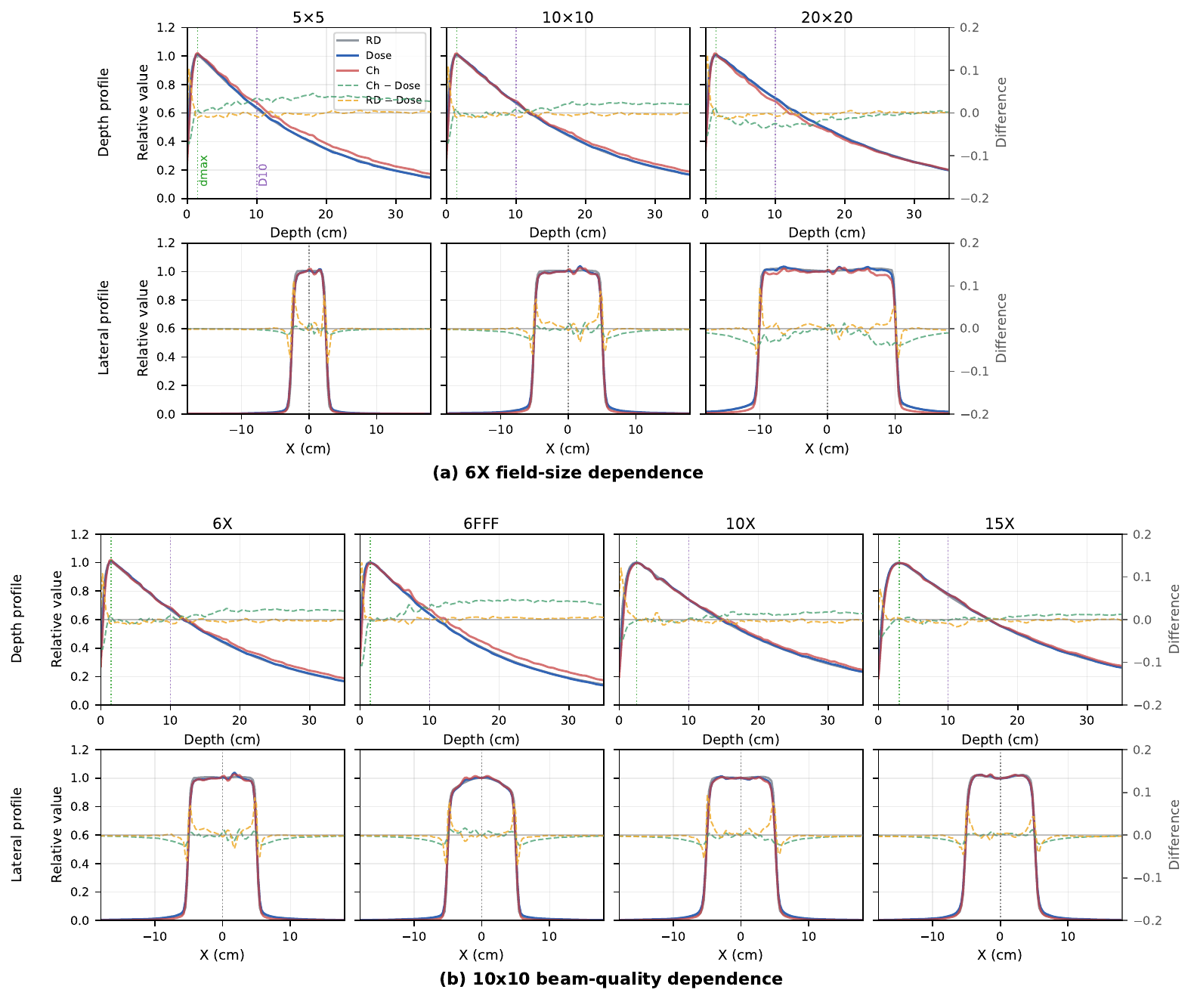}
\captionv{15}{Profile comparisons across field size and photon beam quality}{Profile comparisons across field size and photon beam quality.
Panel (a) shows the 6X  field-size series from \(5\times5\) to \(20\times20\), and panel (b) compares \(10\times10\) fields for 6X, 6FFF, 10X, and 15X.
In each panel, the upper row shows the RD and  dose PDD profiles together with the corresponding normalized Cherenkov profile, and the lower row shows lateral X profiles at the RD-defined \(d_{\mathrm{max}}\).
Solid curves show RD,  dose, and  Cherenkov. In the legends, Ch denotes Cherenkov.
With  dose as the baseline, orange dashed curves show RD minus  dose and green dashed curves show  Cherenkov minus  dose. The right axes show differences between the normalized profiles.
Depth-profile markers denote \(d_{\mathrm{max}}\) and \(D_{10}\), where \(D_{10}\) means the central-axis location at 10 cm depth, not a dose level.
\label{fig:dose_profiles_fixedfield_energy}}
\end{figure}
The profiles in Figure~\ref{fig:dose_profiles_fixedfield_energy}(a) provide a one-dimensional view of the features observed alongside the plane-level and full-volume field-size results.
All three fields retained similar buildup behavior and a gradual dose decrease beyond \(d_{\mathrm{max}}\), while the lateral profiles showed the expected widening of the central plateau and outward movement of the field boundaries.
For \(5\times5\), the  Cherenkov-minus-dose curve showed a sustained separation through much of the depth profile and visible excursions near the lateral field boundaries.
\begin{table}[!htbp]
\centering
\small
\captionv{15}{Lateral  dose penumbra proxy for jaw-defined fixed-field cases}{Lateral  dose penumbra proxy for jaw-defined fixed-field cases.
The field-size rows evaluate 6X jaw-defined square fields, and the beam-quality rows evaluate \(10\times10\) jaw-defined fields with 10X, 15X, and 6FFF beams.
Measurements use the RD-defined \(d_{\mathrm{max}}\), \(Z=0\) lateral X profile.
The final column is the sum of the left and right 80\%--20\% widths divided by \(W_{50}\); it is a one-dimensional geometric proxy, not an area or volume fraction.
\label{tab:penumbra_proxy}}
\setlength{\tabcolsep}{5pt}
\begin{tabular*}{\textwidth}{@{\extracolsep{\fill}}llccc@{}}
\hline
Energy & Modality & \shortstack{Measured field\\width \(W_{50}\) (cm)} &
\shortstack{Mean 80\%--20\%\\penumbra width (cm)} &
\shortstack{Two-edge penumbra\\fraction (\%)} \\
\hline
6X & \(5\times5\)~cm\(^2\) & 5.05 & 0.42 & 16.5 \\
6X & \(10\times10\)~cm\(^2\) & 10.13 & 0.41 & 8.1 \\
6X & \(20\times20\)~cm\(^2\) & 20.29 & 0.44 & 4.3 \\
10X & \(10\times10\)~cm\(^2\) & 10.23 & 0.56 & 10.9 \\
15X & \(10\times10\)~cm\(^2\) & 10.29 & 0.56 & 10.8 \\
6FFF & \(10\times10\)~cm\(^2\) & 10.12 & 0.44 & 8.7 \\
\hline
\end{tabular*}
\end{table}
The 50\% field width \(W_{50}\), mean 80\%--20\% penumbra width \(\bar p_{80-20}\), and two-edge penumbra fraction \(f_{\mathrm{edge}}\) were calculated according to Eqs.~(\ref{eq:w50})--(\ref{eq:edge_fraction}), respectively, and are reported in Table~\ref{tab:penumbra_proxy}.
Among these quantities, \(f_{\mathrm{edge}}\) measures the combined penumbra widths at the left and right field edges relative to \(W_{50}\).
Although the absolute 80\%--20\% penumbra widths were similar across the 6X field-size series at 0.41--0.44~cm, the two-edge fraction decreased from 16.5\% for \(5\times5\) to 8.1\% for \(10\times10\) and 4.3\% for \(20\times20\).
Viewed alongside the gamma and profile findings, the \(5\times5\) case had the lowest full-volume passing rate in the field-size series, differences visible with depth and near the field edges in the lower-dimensional views, and the largest relative edge fraction, \(f_{\mathrm{edge}}=16.5\%\).
Because the absolute penumbra widths were similar across the three fields but \(W_{50}\) was smallest for \(5\times5\), the two edge regions occupied a larger fraction of its lateral field width than in the \(10\times10\) and \(20\times20\) cases.
\FloatBarrier
The beam-quality series held the aperture at \(10\times10\) and compared the spatial patterns for 6X, 6FFF, 10X, and 15X.
In the full-volume analysis, 6FFF had a markedly lower  Dose--Cherenkov \(P_{\gamma}^{(3)}\) of 32.82\%, whereas the other three beam qualities remained above 97.9\%.
The spatial features accompanying this contrast were then examined using profiles and representative planes.
The measured field widths and two-edge fractions remained similar across the beam-quality cases, with \(f_{\mathrm{edge}}\) ranging from 8.1\% to 10.9\% (Table~\ref{tab:penumbra_proxy}).
The 6FFF full-volume reduction was not accompanied by a larger relative edge fraction.
The profiles in Figure~\ref{fig:dose_profiles_fixedfield_energy}(b) retained a common lateral field width, while 6FFF showed the largest sustained  Cherenkov-minus-dose separation along depth.
The representative-plane maps then provided the corresponding two-dimensional view of this depth-profile pattern.
\begin{figure}[H]
\centering
\includegraphics[width=0.99\textwidth]{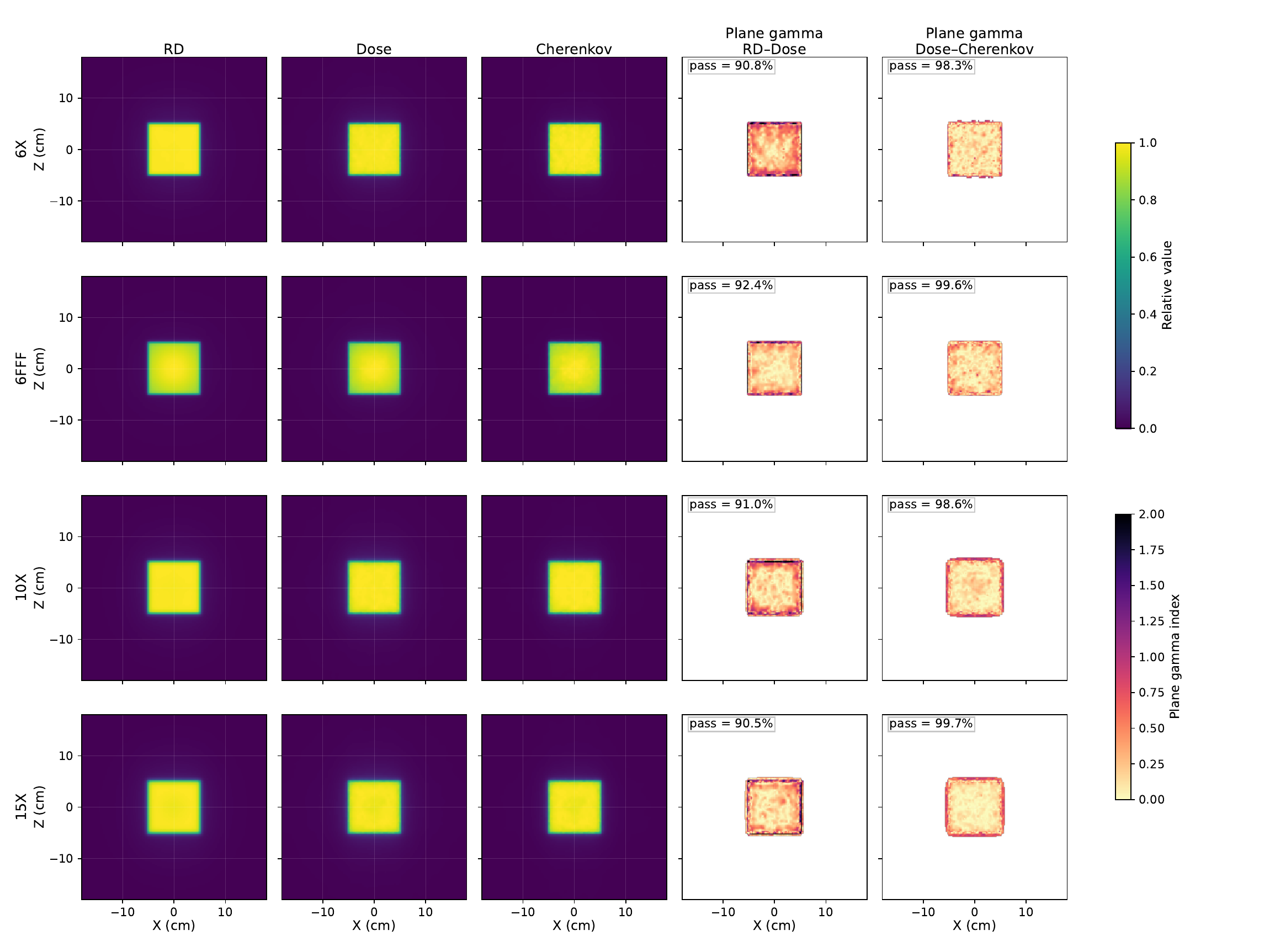}
\captionv{15}{X--Z maps at the RD-defined dmax for 10x10 fields across photon beam qualities}{X--Z maps at the RD-defined \(d_{\mathrm{max}}\) for \(10\times10\) fields across photon beam qualities.
Rows show 6X, 6FFF, 10X, and 15X using the same jaw-defined aperture.
Columns show RD,  dose,  Cherenkov creation counts, the representative-plane 2D  RD--Dose gamma map, and the representative-plane 2D  Dose--Cherenkov gamma map.
The in-panel pass labels report \(P_{\gamma}^{(2)}\) from the independent 2D calculations on these transverse planes.
\label{fig:dose_gamma_energy_xz_cherenkov}}
\end{figure}
Figure~\ref{fig:dose_gamma_energy_xz_cherenkov} shows that all four beam qualities retained the common \(10\times10\) lateral opening on the transverse X--Z plane at the RD-defined \(d_{\mathrm{max}}\).
The  RD--Dose gamma maps  showed similar passing rates across the four beam qualities.
The corresponding  Dose--Cherenkov values were 98.3\%, 99.6\%, 98.6\%, and 99.7\% for 6X, 6FFF, 10X, and 15X, respectively.
The 6FFF transverse plane retained  high  Dose--Cherenkov agreement in this series despite having the lowest full-volume  Dose--Cherenkov result.
\begin{figure}[H]
\centering
\includegraphics[width=0.99\textwidth]{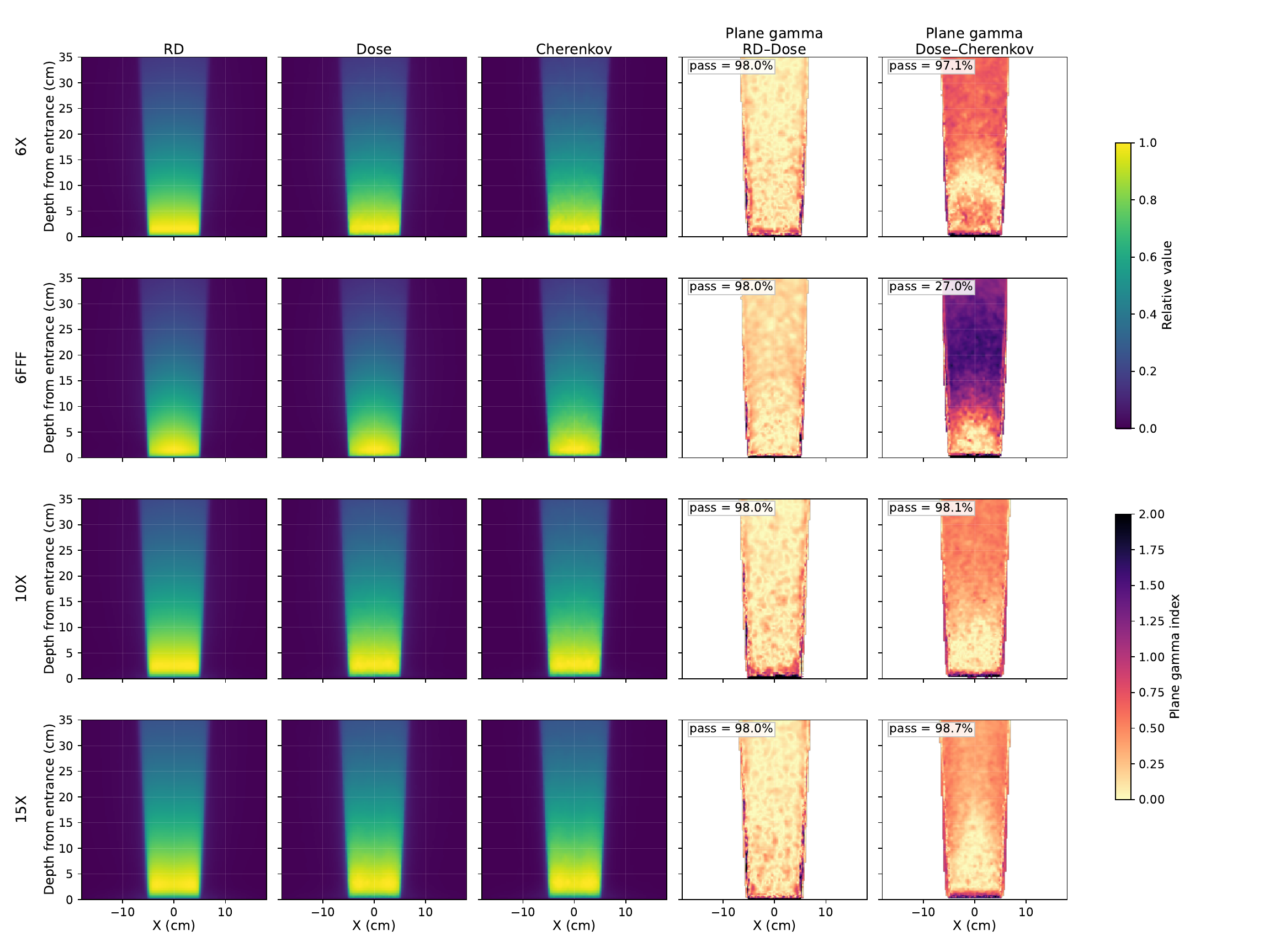}
\captionv{15}{Central X--Y maps for 10x10 fields with different photon beam qualities}{Central X--Y maps for \(10\times10\) fields with different photon beam qualities.
The central \(Z=0\) X--Y plane is longitudinal: Y is the beam-depth direction and is displayed as depth from the phantom entrance.
Rows show 6X, 6FFF, 10X, and 15X using the same jaw-defined aperture.
Columns show RD,  dose,  Cherenkov creation counts, the representative-plane 2D  RD--Dose gamma map, and the representative-plane 2D  Dose--Cherenkov gamma map.
In-panel pass labels report \(P_{\gamma}^{(2)}\) for this independent longitudinal-plane calculation.
\label{fig:dose_gamma_energy_xy_cherenkov}}
\end{figure}
Figure~\ref{fig:dose_gamma_energy_xy_cherenkov} shows a sharp contrast between the longitudinal X--Y results and the transverse-plane results.
 RD--Dose \(P_{\gamma}^{(2)}\) rounded to 98.0\% for every beam quality, whereas  Dose--Cherenkov \(P_{\gamma}^{(2)}\) was 97.1\%, 27.0\%, 98.1\%, and 98.7\% for 6X, 6FFF, 10X, and 15X, respectively.
Only the 6FFF row showed gamma-index values that increased beyond the near-entrance region and remained elevated over much of the subsequent beam depth, forming a broad depth-oriented low-agreement region.
The low full-volume 6FFF agreement coincided with a 27.0\% longitudinal-plane passing rate and sustained depth-profile separation, whereas its transverse-plane agreement remained high at 99.6\%.
Across the jaw-defined cases, lower full-volume  Dose--Cherenkov agreement was accompanied by different patterns in the representative planes and profiles.
The \(5\times5\) field combined edge-visible profile differences, increasing longitudinal gamma-index values with depth, and the largest relative edge fraction.
The \(20\times20\) difference was more prominent on the transverse plane, whereas the 6FFF observations included sustained depth-profile separation and a broad longitudinal low-agreement region despite high transverse agreement.
The VMAT analysis next examined whether high overall association and localized  Dose--Cherenkov differences persisted after spatial modulation and weighted plan composition.
\FloatBarrier
\subsubsection{\texorpdfstring{Dose--Cherenkov relationship in the VMAT delivery}{Dose--Cherenkov relationship in the VMAT delivery}}
Before evaluating the merged two-arc composite, a separate control-point-level analysis was performed to demonstrate the framework's ability to instantiate and simulate complex MLC-shaped delivery states extracted from the modulated plan.
For this analysis, one representative control-point-defined state was selected from each arc and executed as an independent simulation using its RTPLAN-derived aperture and collimator geometry.
The two standalone simulations were analyzed separately from the plan-level composite.
\begin{figure}[!t]
\centering
\makebox[\textwidth][c]{
\includegraphics[width=1.035\textwidth]{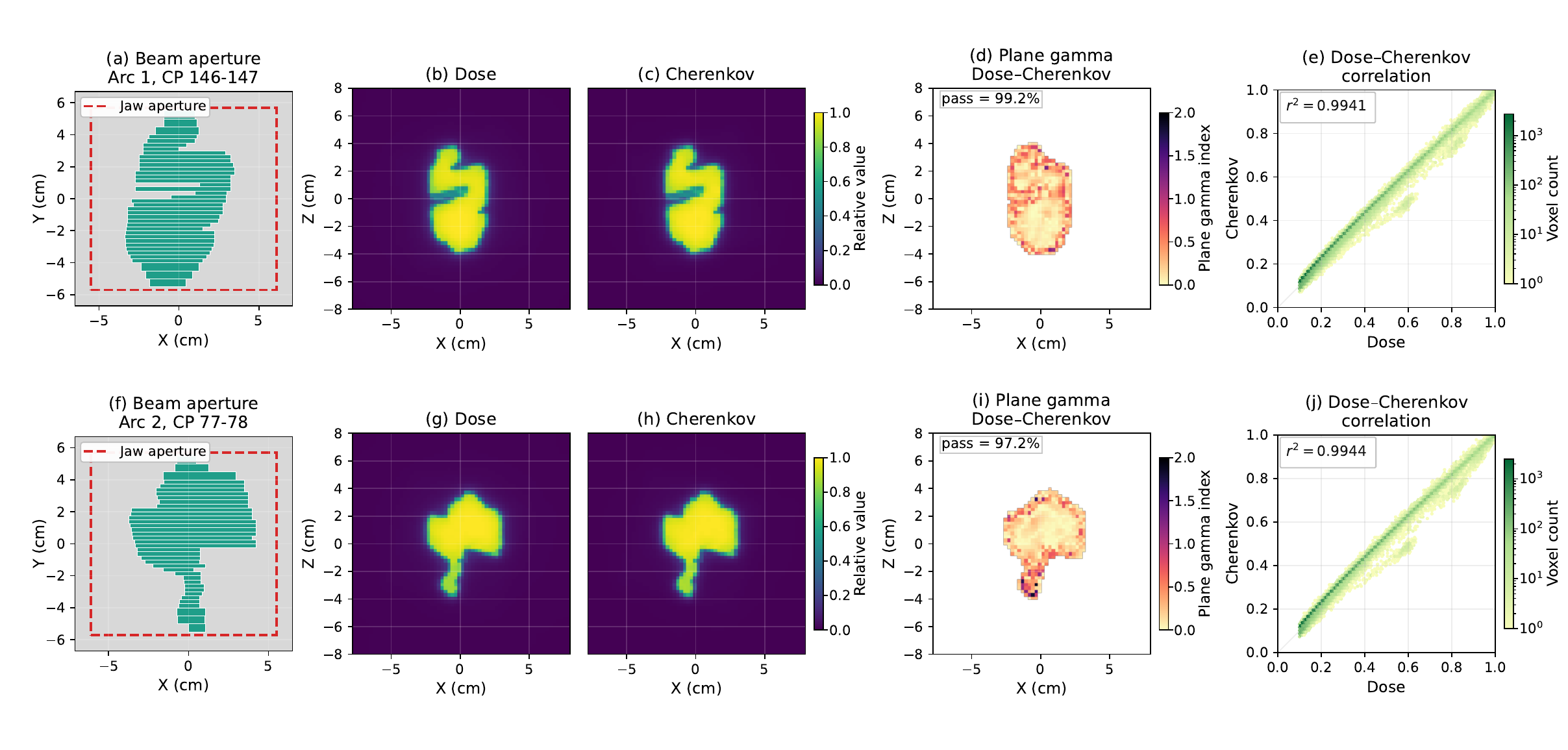}
}
\captionv{15}{Representative VMAT delivery-state diagnostics}{Representative VMAT delivery-state diagnostics.
The top and bottom rows show one control-point-defined delivery state from each arc in a common gantry-zero diagnostic orientation using the complete merged 6X phase-space source; the two selected states were extracted from the RTPLAN and run as separate representative delivery-state simulations.
Each state retains its RTPLAN-derived MLC, jaw, and collimator geometry.
Panels (a,f) show the MLC opening within the jaw-defined outer aperture at isocenter.
Panels (b,g) and (c,h) show the corresponding  dose and  Cherenkov creation distributions on the state-specific representative X--Z plane selected through that delivery state's  dose global-maximum ROI center.
Panels (d,i) show the independently calculated representative-plane 2D  Dose--Cherenkov gamma maps using a 10\% threshold relative to the maximum of the reference plane, with in-panel labels reporting \(P_{\gamma}^{(2)}\); these maps are not slices of a 3D gamma result.
Panels (e,j) show the voxelwise 3D  Dose--Cherenkov correlations evaluated from the paired three-dimensional distributions. The displayed correlation axes are limited to 0--1; the reported \(r^2\) values include all selected voxels.
All quantities within a row come from the same representative delivery state.
\label{fig:vmat_control_point_examples}}
 \end{figure}
\FloatBarrier
Figure~\ref{fig:vmat_control_point_examples} presents the paired  dose and Cherenkov records generated by these two isolated simulations.
The representative-plane  Dose--Cherenkov \(P_{\gamma}^{(2)}\) values were 99.2\% for the selected Arc~1 state and 97.2\% for the selected Arc~2 state.
A separate voxelwise 3D  Dose--Cherenkov correlation was also calculated from the paired volume produced by each isolated simulation, yielding \(r^2=0.9941\) and 0.9944, respectively.
Both isolated control-point simulations retained high voxelwise association and high representative-plane  Dose--Cherenkov agreement.
The merged two-arc composite was then evaluated as the plan-level VMAT result.
\begin{figure}[!htbp]
\centering
\includegraphics[width=0.99\textwidth]{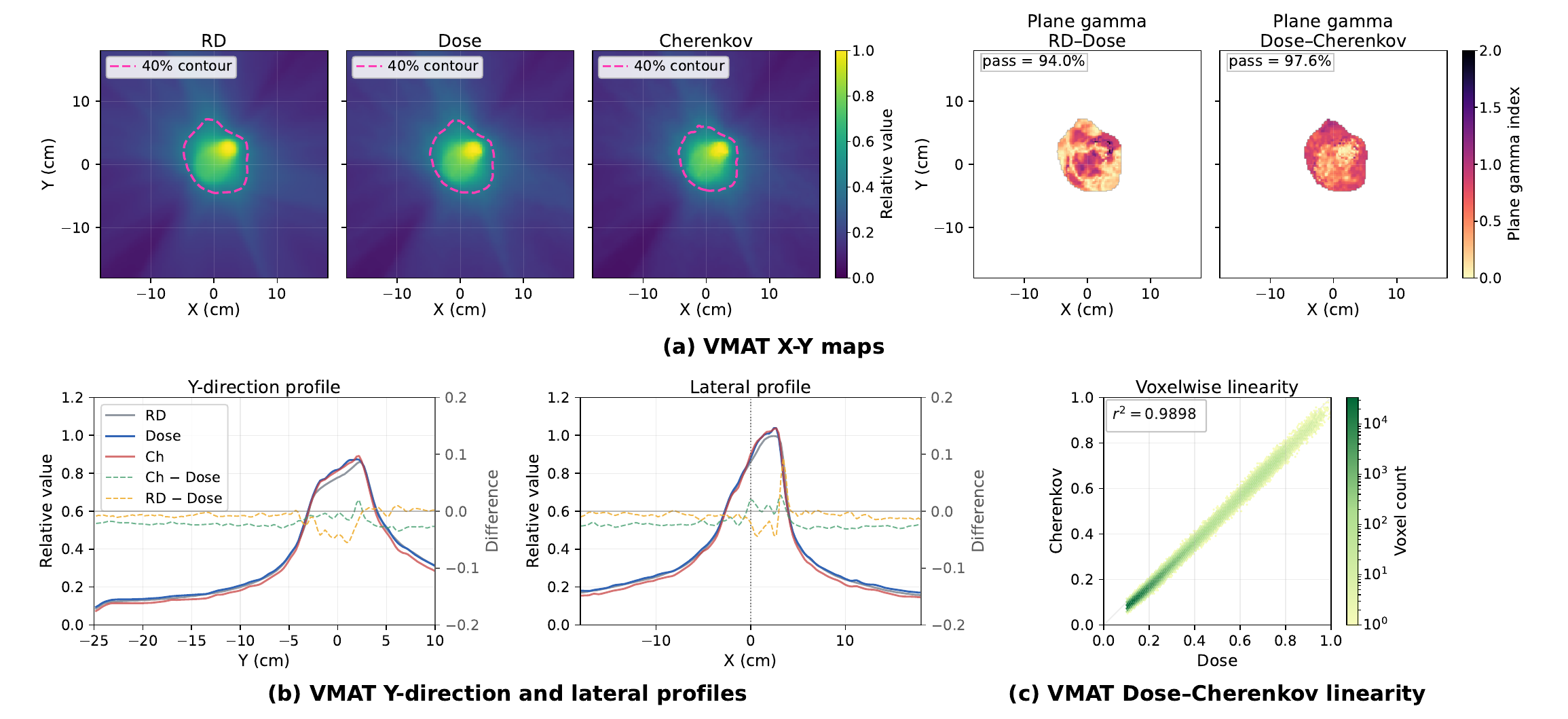}
\captionv{15}{VMAT X--Y maps, profiles, and linearity}{VMAT X--Y maps, profiles, and linearity.
Panel (a) shows the representative X--Y plane through the RD global-3D-maximum ROI center for the final plan-level VMAT composite.
Columns show RD,  dose,  Cherenkov creation counts, representative-plane 2D  RD--Dose gamma, and representative-plane 2D  Dose--Cherenkov gamma.
In-panel pass labels report the independently calculated \(P_{\gamma}^{(2)}\) values of 94.0\% and 97.6\% for  RD--Dose and  Dose--Cherenkov, respectively.
Each representative-plane gamma calculation uses a 40\% reference threshold, and each magenta dashed contour marks the corresponding map's 40\% level.
Panel (b) shows the  Y-direction and lateral X profiles; the 40\% threshold is not applied to these profiles. In the legends, Ch denotes Cherenkov. The right axes show differences between the normalized profiles.
Panel (c) shows the full-volume 3D voxelwise  Dose--Cherenkov linearity plot.
\label{fig:vmat_arc_gamma_validation}}
\end{figure}
\FloatBarrier
Figure~\ref{fig:vmat_arc_gamma_validation} provides an integrated plan-level view of the composite. In contrast to the 10\% threshold used for the representative planes in Section~\ref{sec:results_fieldsize_beamquality_planes}, the VMAT plane in panel (a) used a 40\% threshold.
In panel (b), the Y-direction profiles were extracted at \(X=0\) and \(Z=0\), and the lateral X profiles were extracted at  the Y coordinate of the RD profile maximum and \(Z=0\).
Solid curves show RD,  dose, and  Cherenkov, and dashed curves show  Cherenkov minus  dose and RD minus  dose on the right axes.
The  Y-direction profiles share a common broad rise and peak, while the dashed difference curves show that the residual separation varies  along Y rather than remaining constant.
The lateral profiles retain a common centered peak, while the dashed difference curves identify localized separation across the central peak and its high-gradient sides.
Panel (c) uses the Figure~\ref{fig:cherenkov_voxel_correlation}-style hexbin density representation for the paired full-volume distributions.
The full-volume 3D voxelwise  Dose--Cherenkov correlation yielded \(r^2=0.9898\).
The X--Z projection and  Y-direction profile  Dose--Cherenkov correlations were higher, with \(r^2=0.9987\) and 0.9976, respectively (Table~\ref{tab:cherenkov_full_correlations}).
Together with the representative-plane and profile results, these correlations indicate high overall association with direction-dependent residual separation.
Across the jaw-defined and VMAT deliveries, local Cherenkov creation retained the broad spatial organization of the paired  dose, while the residual  Dose--Cherenkov differences appeared in case-specific patterns at field edges, along beam depth, and within the VMAT composite distribution.
\section{Discussion}

\subsection{Framework reliability from RT-derived delivery translation}

The  RD--Dose results support the dose-producing chain across the tested delivery conditions.
Across jaw-defined fixed fields, beam-quality substitutions, and VMAT composition, the comparisons support the reliability of the RTPLAN-derived delivery translation and voxelized dose-scoring chain within the tested water-phantom setting.
The VMAT composite provided the most demanding plan-level case in the present study because the aperture and weighting were determined by RTPLAN control-point states rather than by a single static jaw opening.

The additional AAA-based analysis provided a secondary TPS-reference comparison for the 6X \(10\times10\) case. Despite the different dose-calculation formulations described in Section~II.B, both the AXB- and AAA-referenced comparisons showed high spatial agreement between dose and the corresponding RD.

\subsection{Dose--Cherenkov spatial relationship in paired records}

As reported in Sections~\ref{sec:results_correlation} and~\ref{sec:results_gamma}, dose and Cherenkov creation remained strongly correlated across the evaluated cases, whereas the full-volume Dose--Cherenkov 3D gamma passing rates varied considerably.
Understanding this contrast requires examining where the two distributions agree and where they diverge.
The plane and profile analyses showed that their differences varied with depth and across the field.

The reductions in full-volume  Dose--Cherenkov agreement were most pronounced for the \(5\times5\) and 6FFF cases, with passing rates of 71.22\% and 32.82\%, respectively; at the large-field end, the \(20\times20\) result of 90.79\% was also lower than the 97.97\% value for \(10\times10\).
To locate the differences underlying the full-volume Dose--Cherenkov results, we examined the representative transverse and longitudinal planes.
Across the field-size and beam-quality series, longitudinal-plane passing rates exceeded 96\% except for the \(5\times5\) and 6FFF cases. Transverse-plane passing rates exceeded 98\% for all beam-quality cases, whereas the \(5\times5\) and \(20\times20\) cases in the field-size series showed lower agreement.

 The depth and lateral profiles provided a more detailed view of how these differences varied within the planes.
For \(5\times5\) and 6FFF, the normalized  Cherenkov curve increasingly separated from the  dose curve with depth and remained distinct over the deeper region.
For \(20\times20\), in contrast, the normalized  Cherenkov curve fell below the  dose curve over shallow-to-intermediate depths, reached its greatest separation while the relative dose remained high, and then moved back toward the  dose curve at greater depth.
These distinct profiles indicate case-dependent spatial differences rather than a common depth-profile distortion underlying all of the gamma reductions.

Depth-dependent spectral evolution provides one context for these profile differences.
Beyond \(d_{\mathrm{max}}\), preferential attenuation of lower-energy photons hardens a polyenergetic beam and shifts the secondary-electron spectrum toward higher energies.
Because Cherenkov creation depends on electron track length above the Cherenkov threshold, this spectral shift can increase Cherenkov creation per deposited dose and make the normalized Cherenkov profile decrease more slowly than the  dose profile.\cite{Glaser2014CherenkovDose,Glaser2014VideoRateCherenkov}
The increasing depth-profile separation observed for \(5\times5\) and 6FFF is consistent with this expected direction.
For 6FFF, removal of the flattening filter also changes the initial beam quality and reduces head scatter relative to flattened 6X, providing additional context for its sustained depth separation.\cite{Feng2016TrueBeamFFFSmallFields}

The \(10\times10\) curves remained closely aligned, whereas the \(20\times20\) field showed the distinct below-then-recover depth pattern described above.
Progressive beam hardening alone does not account for the initial decrease in the normalized Cherenkov-to-dose relationship in this larger field.
Larger fields receive a greater contribution from low-energy photons scattered within the phantom and can consequently exhibit a softer in-field spectrum; this field-size dependence has been demonstrated for 6-MV beams, including direct comparison of \(5\times5\), \(10\times10\), and \(20\times20\) fields.\cite{Scarboro2011PhotonSpectra}
The \(20\times20\) profile is therefore consistent with competition between scatter-induced spectral softening over shallow-to-intermediate depths and progressive beam hardening at greater depth.

In the lateral direction, the large two-edge penumbra fraction for \(5\times5\) and the edge-localized differences in the representative planes identify the field boundary as an additional region of sensitivity.
Charged-particle-equilibrium theory is relevant to this boundary context because the lateral range of secondary electrons determines when electron-transport zones from opposing field edges overlap as field size decreases.\cite{Das2008SmallFieldNonequilibrium}
The larger relative penumbra contribution in the \(5\times5\) field is consistent with this boundary-sensitive transport picture and with the observed concentration of  Dose--Cherenkov differences near the field edges.

The VMAT analysis extended the paired-record evaluation to a modulated delivery.
The standalone control-point simulations demonstrated execution of selected MLC-shaped delivery states, while the merged two-arc composite retained high dose-related association and a full-volume  Dose--Cherenkov gamma passing rate of 96.59\%.
The framework generated paired records across control-point-dependent MLC apertures, gantry and collimator states, meterset weighting, and plan-level composition.

\subsection{Limitations and future validation}

The present study focuses on validating the paired-record generation workflow in a homogeneous water phantom.
It does not yet address optical transport, patient anatomy, geometric complexity, or emission boundary conditions.
Future studies should extend the same paired-record workflow to heterogeneous and patient-like geometries, where CT material assignment and anatomical boundaries can affect dose deposition and charged-particle transport, while refractive-index and optical-property choices influence Cherenkov creation, transport, and escape.

The Cherenkov outputs reported here are internal photon-creation records, not camera images.
This source-record level is useful for paired-data generation because it preserves the local relationship between charged-particle transport, dose deposition, and Cherenkov creation before optical transport and detector effects are introduced.
Applications involving measured Cherenkov images will require additional layers for optical transport, boundary escape, camera response, calibration, and comparison with phantom or clinical measurements.

The current study focuses on one mainstream linac configuration (Varian TrueBeam with an HD120 MLC).
The  RD--Dose comparison in this manuscript is a relative spatial-dose reference for the data-generation framework.
Further validation could incorporate measured beam data, machine-specific beam modeling, MLC characterization, absolute calibration, and more complex benchmark plans with target and organ-at-risk geometries.
These extensions would move the framework toward more complete structure-aware and measurement-referenced validation.
The contribution demonstrated here is narrower and upstream: a traceable RT-derived Monte Carlo workflow for generating paired  dose-deposition and Cherenkov photon-creation records for controlled analysis and future dataset construction.

\section{Conclusion}

This work demonstrates a precise Geant4-based digital linac framework for generating paired  dose and Cherenkov distributions from RT datasets.
The framework translates RT dataset information into executable fixed-field and VMAT Monte Carlo deliveries, computes dose and Cherenkov quantities under the same delivery and phantom conditions, and preserves the provenance needed to connect the resulting distributions to their source, delivery, and scoring settings.
Across the evaluated  field-size, beam-quality, and VMAT cases, the  dose distributions showed spatial consistency with TPS-computed RD, while the paired Cherenkov distributions maintained strong dose-related spatial association.
These results support the framework as a controlled generator of paired  dose--Cherenkov-creation datasets, providing a foundation for quantitative analysis of the relationship between local Cherenkov creation and deposited dose.

The present distribution-level framework also defines the starting point for the next stage of Cherenkov modeling.
Future work will build on these paired dose and Cherenkov distributions by adding tissue-optics-based photon transport in RT volumes, estimating the Cherenkov signal that escapes to the patient or phantom surface, and correlating the predicted surface-emitted signal with superficial dose measurements.
This extension will connect the controlled internal distributions generated here with the surface signals used in Cherenkov-based treatment monitoring.

\section*{Author contributions}

Xianghuan He performed the experiments, analyzed the data, and drafted the manuscript. Changyu Sun, Andy N. Wiles, Yao Hao, Gregory B. Biedermann, Brian W. Pogue, and Rongxiao Zhang reviewed and revised the manuscript. All authors contributed to scientific discussions throughout the study.

\section*{Conflict of interest statement}

The authors have no relevant conflicts of interest to disclose.

\section*{Data and code availability}

The simulation data and custom simulation and analysis code supporting this study are available from the corresponding author upon reasonable request. Access to and redistribution of third-party phase-space files and treatment-planning inputs are subject to the relevant providers' conditions.

\clearpage
\section*{Figure captions}

\begin{samepage}
\noindent\textbf{Figure~\ref{fig:overview}.} Overview of the DICOM RT-driven framework for generating paired  dose and local Cherenkov-creation records.
The four stages are input data, delivery geometry construction from RTPLAN control point (CP) information, Monte Carlo simulation, and records and provenance.
Beam-source data, the computed tomography (CT)-derived phantom, and RTPLAN define each simulated case.

Energy deposition and Cherenkov photon creation positions are scored on a common voxel grid within the same transport history.
TPS-computed RTDOSE remains external to particle transport and is used only as the reference for  RD--Dose comparison.
\par
\end{samepage}
\medskip

\begin{samepage}
\noindent\textbf{Figure~\ref{fig:cherenkov_voxel_correlation}.} Voxelwise 3D  Dose--Cherenkov correlations.
Hexbin density plots compare paired  dose and local Cherenkov creation intensity on the common analysis grid, with each axis independently rescaled by the maximum of its corresponding 3D record.
The \(10\times10\) beam-quality comparison spans 6X, 6FFF, 10X, and 15X, with the 6X case shown as the middle panel of the bottom row.
The bottom row fixes the photon beam quality at 6X and compares \(5\times5\), \(10\times10\), and \(20\times20\) fields.
Color indicates voxel count per hexagonal bin on a logarithmic scale, with a separate count scale for each row.
The \(r^2\) values quantify association among the paired voxel values; color intensity and the faint diagonal guide do not define the correlation or agreement with a reference line.
\par
\end{samepage}
\medskip

\begin{samepage}
\noindent\textbf{Figure~\ref{fig:dose_gamma_fixedfield_xz_cherenkov}.} X--Z maps at the RD-defined \(d_{\mathrm{max}}\) for three 6X jaw apertures.
Rows show \(5\times5\), \(10\times10\), and \(20\times20\) fields generated from the RTPLAN X and Y jaw settings.
Columns show RD,  dose,  Cherenkov creation counts, the representative-plane 2D  RD--Dose gamma map, and the representative-plane 2D  Dose--Cherenkov gamma map. The relative-value colorbar applies to the first three columns, and the gamma-index colorbar to the last two.
The in-panel pass labels report \(P_{\gamma}^{(2)}\) from these independent 2D calculations; they do not report a slice of the full-volume 3D result.
\par
\end{samepage}
\medskip

\begin{samepage}
\noindent\textbf{Figure~\ref{fig:dose_gamma_fixedfield_xy_cherenkov}.} Central X--Y maps for three 6X jaw-defined field sizes.
Here, Y is the beam-depth direction and is displayed as depth from the phantom entrance.
Rows show \(5\times5\), \(10\times10\), and \(20\times20\) fields on the \(Z=0\) plane.
Columns match Figure~\ref{fig:dose_gamma_fixedfield_xz_cherenkov}: RD,  dose,  Cherenkov creation counts, the representative-plane 2D  RD--Dose gamma map, and the representative-plane 2D  Dose--Cherenkov gamma map.
In-panel pass labels report \(P_{\gamma}^{(2)}\) for this longitudinal plane, where Y is depth.
\par
\end{samepage}
\medskip

\begin{samepage}
\noindent\textbf{Figure~\ref{fig:dose_profiles_fixedfield_energy}.} Profile comparisons across field size and photon beam quality.
Panel (a) shows the 6X  field-size series from \(5\times5\) to \(20\times20\), and panel (b) compares \(10\times10\) fields for 6X, 6FFF, 10X, and 15X.
In each panel, the upper row shows the RD and  dose PDD profiles together with the corresponding normalized Cherenkov profile, and the lower row shows lateral X profiles at the RD-defined \(d_{\mathrm{max}}\).
Solid curves show RD,  dose, and  Cherenkov. In the legends, Ch denotes Cherenkov.
With  dose as the baseline, orange dashed curves show RD minus  dose and green dashed curves show  Cherenkov minus  dose. The right axes show differences between the normalized profiles.
Depth-profile markers denote \(d_{\mathrm{max}}\) and \(D_{10}\), where \(D_{10}\) means the central-axis location at 10 cm depth, not a dose level.
\par
\end{samepage}
\medskip

\begin{samepage}
\noindent\textbf{Figure~\ref{fig:dose_gamma_energy_xz_cherenkov}.} X--Z maps at the RD-defined \(d_{\mathrm{max}}\) for \(10\times10\) fields across photon beam qualities.
Rows show 6X, 6FFF, 10X, and 15X using the same jaw-defined aperture.
Columns show RD,  dose,  Cherenkov creation counts, the representative-plane 2D  RD--Dose gamma map, and the representative-plane 2D  Dose--Cherenkov gamma map.
The in-panel pass labels report \(P_{\gamma}^{(2)}\) from the independent 2D calculations on these transverse planes.
\par
\end{samepage}
\medskip

\begin{samepage}
\noindent\textbf{Figure~\ref{fig:dose_gamma_energy_xy_cherenkov}.} Central X--Y maps for \(10\times10\) fields with different photon beam qualities.
The central \(Z=0\) X--Y plane is longitudinal: Y is the beam-depth direction and is displayed as depth from the phantom entrance.
Rows show 6X, 6FFF, 10X, and 15X using the same jaw-defined aperture.
Columns show RD,  dose,  Cherenkov creation counts, the representative-plane 2D  RD--Dose gamma map, and the representative-plane 2D  Dose--Cherenkov gamma map.
In-panel pass labels report \(P_{\gamma}^{(2)}\) for this independent longitudinal-plane calculation.
\par
\end{samepage}
\medskip

\begin{samepage}
\noindent\textbf{Figure~\ref{fig:vmat_control_point_examples}.} Representative VMAT delivery-state diagnostics.
The top and bottom rows show one control-point-defined delivery state from each arc in a common gantry-zero diagnostic orientation using the complete merged 6X phase-space source; the two selected states were extracted from the RTPLAN and run as separate representative delivery-state simulations.
Each state retains its RTPLAN-derived MLC, jaw, and collimator geometry.
Panels (a,f) show the MLC opening within the jaw-defined outer aperture at isocenter.
Panels (b,g) and (c,h) show the corresponding  dose and  Cherenkov creation distributions on the state-specific representative X--Z plane selected through that delivery state's  dose global-maximum ROI center.
Panels (d,i) show the independently calculated representative-plane 2D  Dose--Cherenkov gamma maps using a 10\% threshold relative to the maximum of the reference plane, with in-panel labels reporting \(P_{\gamma}^{(2)}\); these maps are not slices of a 3D gamma result.
Panels (e,j) show the voxelwise 3D  Dose--Cherenkov correlations evaluated from the paired three-dimensional distributions. The displayed correlation axes are limited to 0--1; the reported \(r^2\) values include all selected voxels.
All quantities within a row come from the same representative delivery state.
\par
\end{samepage}
\medskip

\begin{samepage}
\noindent\textbf{Figure~\ref{fig:vmat_arc_gamma_validation}.} VMAT X--Y maps, profiles, and linearity.
Panel (a) shows the representative X--Y plane through the RD global-3D-maximum ROI center for the final plan-level VMAT composite.
Columns show RD,  dose,  Cherenkov creation counts, representative-plane 2D  RD--Dose gamma, and representative-plane 2D  Dose--Cherenkov gamma.
In-panel pass labels report the independently calculated \(P_{\gamma}^{(2)}\) values of 94.0\% and 97.6\% for  RD--Dose and  Dose--Cherenkov, respectively.
Each representative-plane gamma calculation uses a 40\% reference threshold, and each magenta dashed contour marks the corresponding map's 40\% level.
Panel (b) shows the  Y-direction and lateral X profiles; the 40\% threshold is not applied to these profiles. In the legends, Ch denotes Cherenkov. The right axes show differences between the normalized profiles.
Panel (c) shows the full-volume 3D voxelwise  Dose--Cherenkov linearity plot.
\par
\end{samepage}
\medskip

\end{document}